# DIGITAL RESILIENCE TO COVID-19: A MODEL FOR NATIONAL DIGITAL HEALTH SYSTEMS TO BOUNCE FORWARD FROM THE SHOCK OF A GLOBAL PANDEMIC

Scott Russpatrick, University of Oslo, scottmr@ifi.uio.no

Johan Sæbø, University of Oslo, johansa@ifi.uio.no

Eric Monteiro, NTNU, eric.monteiro@ntnu.no

Brian Nicholson, University of Manchester, brian.nicholson@manchester.ac.uk

Terje Sanner, University of Oslo, terjeasa@ifi.uio.no

**Abstract:** COVID-19 represented a major shock to global health systems, not the least to resource-challenged regions in the Global South. We report on a case of digital, information system resilience in the response to data needs from the COVID-19 pandemic in two countries in the Global South. In contrast to dominant perspectives where digital resilience enables bounce back or maintenance of a status quo, we identify five *bounce forward* resilience preconditions (i) distributed training, (ii) local expertise (iii) local autonomy and ownership (iv) local infrastructure and (v) platform design infrastructure. These preconditions enable an elevated degree of resilience that in the face of an external shock such as COVID-19 can deliver a bounce forward or strengthening of the information system beyond its pre-shock state.

**Keywords**: COVID-19, Resilience, Bounce forward, ICT4D, Information Systems, Models, Adapt, Change.

## 1. INTRODUCTION

COVID-19 is an exogenous shock disrupting health care systems, business, economies, societies across the entire world [1]. The United Nations defines an exogenous shock as an "unexpected or unpredictable events that occur outside an industry or country but can have a dramatic effect on the performance or markets within an industry or country." [2]. Minimizing the population impact and maximizing vaccine update of COVID-19 is dependent on national disease surveillance systems [3]. Considering the increased focus and dependency on disease surveillance information systems, there is a need to further understand their digital resilience [4]. A helpful starting point is Heeks and Ospina's definition of digital resilience in the context of ICT4D as "the ability of a system to withstand, recover from, and adapt to short-term shocks and longer-term change" [5].

The development agenda, however, calls for information systems to grow and strengthen rather maintain a static state [5]. Marais points out that there is a real concern that resilience of information systems could equate to a stagnation or an approach toward preservation of the status quo [6]. Observed from the prior shocks Ebola and the great East Japan earthquake of 201, Sakuai and Chughtai state that, "a recovery as returning to the pre-disaster state was not enough; resilience requires going beyond rebound, and must encourage adapting to the existing crisis and then transforming." [7]. Therefore, models and drivers of resilience that results in change needs to be explored. Here we refer to the phenomenon of an information system strengthening, improving, or developing in response to an exogenous shock as a "bounce forward". In essence it is a progression past the pre-shock state to a stronger, more developed, and/or scaled state [5].

We examine the use of the free and open source District Health Information System 2 (DHIS2) platform for COVID-19 surveillance. Some 67 mainly low- and middle-income countries have adopted DHIS2 as their central digital platform for their health system [8]. Within the comprehensive DHIS2 ecosystem, we selectively zoom in on the efforts of two countries' Covid 19 responses, viz. Sri Lanka and Sierra Leone. These two countries are sampled for their ability to respond to the pandemic beyond mere "recovering".

In this research we propose the following research question: *What are the enabling preconditions of a resilient information system to experience a bounce-forward when confronted with an exogenous shock?*

The paper is organised as follows. First, we define our grounded methodology of data coding, identification of the preconditions, and model development. Next, we present a detailed description of each case-study. From there, we present our findings of the preconditions and bounce forward resilience models, and finally we provide a brief discussion and reflection on these findings.

## 2. CONCEPTUALIZING DIGITAL RESILIENCE

This paper views digital resilience through ICT4D perspective, where the long-term notion about resilience is change or adaptation rather than solely the short-term notion which is more simply stability in the face of a change or shock [9, 10]. Through the ICT4D perspective we, view resilience as system properties [5, 11]. Many have introduced the concept of system properties that equate towards resilience al be it by different names; system attributes, qualitative, sub properties, and characteristics [5, 11, 12, 13, 14]. Each of these studies and others introduce resilience attributes which some redundancy but little consolidation [5, 11, 15, 16].

Taking all existing IS literature on resilience into account and drawing from the broader disciplines of ecology, engineering, and psychology, Heeks and Ospina introduce foundational resilience attributes that enable an information system to be able to bounce back or return unaffected by a shock. These are:

- Robustness - "Ability of the system to maintain its characteristics and performance in the face of contextual shocks and fluctuations."
- Self-organization - "Ability of the system to independently rearrange its functions and processes in the face of an external disturbance, without being forced by the influence of other external drivers."
- Learning - "Capacity of the system to generate feedback with which to gain or create knowledge and strengthen skills and capacities necessary to experiment and innovate."

Heeks and Ospina also define attributes that enable these foundational attributes as, redundancy, rapidity, scale, diversity and flexibility, and equality [5].

However, again drawing on an ICT4D perspective and acknowledged by Heeks and Ospina, there is an added need for an understanding of resilience as longer-term change or adaption [5, 7]. To that point, only a handful of model of IS resilience exist, none of which incorporate change or adaption / "bounce forward" [5, 11, 16]. Some argue, that the perspective for additional resilience model creation and theory refinements should best be derived from rich interpretative, empirical investigations which there is a paucity of in the existing IS or ICT4D literature [5, 7, 13]. Schryen argues that analysis of rich empirics should be the primary vehicle that IS researcher use to decipher the complexity and dependencies inherent in such a model [17].

Therefore, this paper responds to three gaps in the IS resilience literature. First, we identify specific system properties, that we refer to as preconditions, that enable a bounce forward of the information system. Secondly, we then present a new theoretical resilience model that illustrates the complexities and the dependencies of the information system to be able to bounce forward from an exogenous shock such as COVID-19. Finally, we supply a rich empirical investigation of two longitudinal case-studies in two countries developing and implementing COVID-19 surveillance information systems [4, 5, 7, 17]

## 3. METHODS AND CASE CONTEXT

The context of this study is the ongoing implementation and use of DHIS2. DHIS2 is a free and open source District Health Information System 2 (DHIS2) platform serving 67 low- and middle-income countries in the Global South. DHIS2 has a core database, a suite of generic applications that cover data entry, analysis, and system administration, and an application programming interface (API) which are developed and maintained by the Health Information System Project (HISP) headquartered at the University of Oslo. Beyond the core is a continuously increasing number of third-party developed applications that are developed with little or no involvement from the core development team in Oslo. These periphery, generic applications are by nature more reusable across countries and contexts and increase the value of the platform as a whole to all users [18]. Our research design is informed by theoretically sampling of two out of 67 countries in the DHIS2 ecosystem that demonstrates vividly the preconditions and strategies of digital resilience beyond "bouncing back".

The methods use grounded theory as it allows for natural theory evolution during the research process [19]. Specifically, a grounded theory based methodology enables the researcher to deductively process the interviews, observations, and other data into categories, or distinct units of meaning, which can then evolve into inductive a-priori theory generation that is grounded in empirical observation [20, 21].

For our data we conducted two longitudinal, interpretive case studies [22]. We followed from their beginning the implementation of DHIS2 for COVID-19 monitoring in Sri Lanka and Sierra Leone. Data was collected through four key informant interviews, text analysis from four publicly available case descriptions, presentations of the use-cases during six webinars, and direct observations from communication and activities between the implementation teams in the country and the implementation support staff based at UiO. Direct observations took place between February 2020 and March 2021. Data from direct observation was gathered from 13 meetings between the UiO COVID-19 implementation team and Sri Lanka and 4 between Sierra Leone and the UiO team. The goal of these meetings were to get an update on the progress of the implementation, identify issues, provide technical guidance, and capture any software issues. Running, detailed notes from these meeting were kept for each case-study. These notes include direct quotes from both UiO and country implementers, status of each implementation at the time of the meeting, future plans, any encountered issues, points of success, etc. Key informants were identified through the communication between UiO implementation support staff and the implementation teams in the country. The first round of semi-structured, key informant interviews was in early October 2020, one informant from both cases. Those key informants were asked who else they recommended be interviewed which led to another round of key informant interviews in late October again with one interview from each case. Each interview was 1-1.5 hours long and transcribed.

We then applied a four-step grounded, interactive process to derive five pre-conditions that resulted in resilience attributes that enabled the information systems to improve or bounce forward from exposure to the shock of COVID-19. First, the data was tabulated into chronological order so that it could be coherently analysed. For example, observation notes were broken into the case study timeline so that quotes from the interviews, webinars, and text from use-case descriptions could be placed into the same chronology to build a coherent sequence of key events. Second, we applied line-by-line open coding to our data to group them into distinct, labelled concepts [23]. Next, we employed axial coding to cluster the concepts to derive the preconditions which were iteratively refined until they offered sufficient explanatory power over the observed phenomenon [20, 21, 24].

Finally, we used selective coding as a reductive approach of refining a theory by identifying 'core' categories which our identified concepts and precondition can be coded against. Selective coding further refines the categories and lays a foundation that additional theory can be built upon [20, 21]. We used the Heeks and Ospina foundational resilience attributes, as previously defined, as our core categories to which we coded concepts and the preconditions [5]. Note that we did not code against the Heeks and Ospina enabling conditions as we pursued to identify novel drives of the bounce forward phenomenon in accordance with our research question. Through selective coding we reductively organized our identified categories into the corresponding resilience attributes. From there, we were able to trace our preconditions to the resilience attributes (figure 2). This approach allowed us to recognize the complex interplay between the pre-conditions, the resilience attributes, the COVID-19 shock, and the observed bounce-forward outcomes thus leading us to a model of bounce-forward information system resilience.

## 4. CASE DESCRIPTION

### 4.1. Sierra Leone

While Sierra Leone has used DHIS2 since 2008, the use of the software for disease surveillance is relatively recent, following the West African ebola virus outbreak. The electronic Case Base Disease Surveillance (eCBS) was set up for both weekly reporting of aggregate cases for 26 diseases, and for handling individual cases for 20-22 of these (malaria for instance would be too many cases to make sense to have case-based). A prominent group of diseases is Acute Respiratory Infections (ARI), which typically give symptoms similar to flu, and includes diseases like SARS and MERS. Patients with acute respiratory syndromes (ARS) will be registered as unconfirmed ARS until a lab confirmation. All these diseases have a quite standardized set of information collected for standardized stages, including lab request, lab result, and split stages depending on the disease. Given the similarity of ARIs, it was a relatively easy task to adjust it to have the additional needed variables for Covid-19. In early February 2020 it took just a couple of days for the lead system architect to set up case-based disease surveillance for COVID-19. The relative knowledge deficiency in the early period of the disease meant that some of the variables were never used, like "have you been to the animal market lately". There have subsequently been minor changes to the variables collected.

The usage of the existing case-based surveillance system was not a given in the early period of the pandemic. A different government directorate was pushing for new technologies and apps. The high level of uncertainty, an international sense of panic, and the strong influence of external actors such as aid agencies, many of whom are supporting their own technologies, also contribute to an environment where new technologies can be seen as more favourable. However, the past experience of the Ebola outbreak was decisive in uniting key actors to build on their current system.

> "One key experience from the ebola epidemic 2014-15 remains in force: Strengthen existing systems during an outbreak!"

> "the last thing you do in a crisis situation is start from scratch" [...]"what was lucky for us, the eCBS had already been institutionalized. If we hadn't been there, there would have been a free-for-all situation", an app for this, an app for that"

Once established as the primary COVID-19 reporting system and quickly configured, the eCBS was able to scale rapidly to the whole country.

> "But the benefit is now that we were only active in four out of 16 districts in January before COVID, February when we started COVID we rolled that rapidly to all districts and now we were then following up with all the other diseases, et cetera, to those districts. So again, we can benefit from all the work that has been done."

Prior to COVID-19 pandemic the eCBS was operating in only four districts, but with the rapid need for COVID-19 surveillance the scale of the system was fast tracked and rapidly scaled to all 16 districts in the country.

> "And as I said, that the benefit here is that all the efforts being put into COVID-19 and setting up the systems and contact tracing, and all of that it's very easy to expand it to the other diseases. You basically almost get that for free. The training now is about is very much anything, pick the detail for each disease exactly, what they data you're capturing, how do you handle each disease, et cetera." [...] "We didn't have capacity to obviously train everybody at the same time, but it was over two or three months."

By rapidly scaling the eCBS driven by the need for COVID-19 surveillance, disease surveillance for all diseases correspondingly scaled to the whole country. The focus became training, which was done incrementally using a training-of-trainers model from February through May. First district health medical technicians were trained then the district staff trained the facility staff. The training includes data capture and surveillance for all diseases, not just COVID-19. In Sierra Leone COVID-19 acted as a catalyst for galvanization around an existing disease surveillance system, rapid configuration of that system to accommodate COVID-19, and the rapid scale of the system expanding all disease surveillance to the whole country. Additionally, a virtual training was provided by external DHIS2 administration experts in March 2020 to the core DHIS2 country administration team on DHIS2 design, management, and database scripting.

The utility of the expanded eCBS has already been tested with a localized outbreak of Ebola in the border regions with Guinea in mid-late February 2021. While the data is very fresh, indications are that the eCBS is performing nominally in this area of concern and providing near real-time information on the current evolving situation.

### 4.2. Sri Lanka

DHIS2 was first introduced in Sri Lanka in 2011 by a group of enthusiastic master's students in the Biomedical Informatics program at the University of Colombo. The first implementations were small for local NGOs, but over time these grew into national program systems owned by the Ministry of Health (MoH). In 2017, in response to the need for more DHIS2 support for the various MoH programs, HISP Sri Lanka was established as a company specializing in supporting DHIS2 implementations throughout the country.

On the 20th of January 2020 top digital health doctors in the MoH met to discuss the need to collect data on and screen travellers arriving to Sri Lanka from areas with a high prevalence of COVID-19. The first case of COVID-19 was reported in Sri Lanka on 27 January 2020. Within two days of that, a DHIS2 implementation group that supports the MoH, HISP Sri Lanka, had developed a new DHIS2 instance to register all travellers arriving into the country through airports.

> "So the major concern was to get this tourist register, then now, then to have follow them up at the level. Because we were kind of very competent in DHIS2, so we could quickly set up something in two to three days time."

The epidemiology unit within the MoH as well as the Director of Health Information Unit urged the Director General of Health Services to swiftly approve the system. With the approval, infrastructure, government database cloud hosting, and hardware were rapidly identified through the government ICT Department. The ICT Department was already experienced hosting DHIS2 instances due to the many that are already deployed in the country. Initially a small team of three DHIS2 experts at HISP Sri Lanka were able to do the inceptive configuration of the information system, but more human resources were needed to implement it in the quarantine centres and ports of entry. This was addressed by drawing on the large, distributed pool of medical doctors what had completed a government sponsored master's in information systems at the University of Colombo. In 2009 the University of Colombo launched a master's in information systems program to train Medical doctors to be Medical Officers in Health Informatics in the Ministry of Health. Over the last decade, this program has produced a large cadre of medical professions that are able to support the digitization and implementation of information systems.

> "So with regard to human resources, we needed people to implement and train the end users. So for that one, we had around another like six other doctors who are already there in the Ministry of Health, some of them who are based in different districts. [...] So of course, the public health staff who at field level, they were kind of familiar with the DHIS2, so training them was kind of not that difficult. But the point of entry staff who were at the airport and the quarantined unit of Ministry of Health, they did not know DHIS2, they had not used DHIS2 before. So for them, the training was like it was a bit time consuming. But that part was mainly taken care of these five or six doctors who were attached to Ministry of Health."

By early February the port-of-entry travel screen system was fully implemented at all airports in Sri Lanka. However, it quickly became apparent that to be able to follow-up with travellers and their contacts throughout the country's existing health infrastructure was required. This necessitated a broader active COVID-19 surveillance system that could record all cases and their contacts across the entire country. Again, HISP Sri Lanka was called on to develop this system.

> "So but what happens is this process things (defining and developing a new system) takes longer time. And this is the usual scenario. But when it comes to COVID, we had to be a child, because now people were also panicking, say, for example, they had their deliverables, like people wanted data, and they wanted some visualizations. And when we try to go for a compromise, this was sometimes not possible (for COVID-19)."

HISP Sri Lanka reached out to the core DHIS2 development leads to enquire if their COVID-19 surveillance requirements were possible within the existing feature sets in DHIS2, which they were not. Specifically, Sri Lanka needed to be able to transfer patients between and from site to site and visualize the index and contact cases in a network graph. HISP Sri Lanka, utilizing DHIS2's open source components and API, decided to build new applications to satisfy these requirements. At the time HISP Sri Lanka did not have the necessary developers internally, so they communicated with the ICT Department which were able to lend a few developers. However, this was still not sufficient to be able to build the applications quickly as was required, so the ICT Department director posted an open invitation for volunteer developers via Twitter and organized a hackathon. Within twenty-four hours twenty-five additional developers were identified. Most of these developers were ethnically Sri Lankan, but they were distributed globally. UiO also made available a lead core developer to assist in the utilization of core components and resources. Starting with a hackathon on March 14-15, in the course of two weeks a new port-of-entry and contact tracing data capture application, a COVID-19 case relationships analytics application, and integration with the Sri Lankan immigration system were developed by this pool of volunteer developers, ICT department developers, HISP Sri Lanka implementation manager, and a UiO core developer working collaboratively.

Prior to COVID-19 Sri Lanka has developed siloed disease or program specific information system. Most of these utilize DHIS2, but there is not a single DHIS2 instance that integrates data across all programs.

> "[...]because of the way we could interact with the stakeholders, and we showed that integrations are possible, we recently received a call from the ministry to have a discussion about setting up a central HMIS so that that is going to be on DHIS2. So we will have a one HMIS with all the dashboards. What we talk about in most of the countries, now, we are seeing that becoming a reality, mainly due to what has happened during the COVID-19 times. So there are good mostly good aspects of whatever we did during that time, which has helped us."

The unilateral focus to implement the COVID-19 surveillance system also spurred deeper connections and collaborations between departments. Prior to the COVID-19 response the Ministry of Health did not have proper access to the ICT Department due to some internal resistance and politics. The ICT Department provides ICT services for all other departments except for the Ministry of Health. During the COVID-19 response the Ministry of Health needed to cooperate with the ICT Department for cloud hosting the scaling COVID-19 surveillance system as well as assistance with developing of the new applications.

> "And now, they're (Ministry of Health) also getting more inputs from the ICT agency about the systems that they're implementing. So that way, the collaborations that really took place during the COVID-19 has actually contributed positively and they are still ongoing."

Sri Lanka was the first country to develop such an extensive surveillance system including new data capture and analytics tools specifically designed for COVID-19. They felt the need to provide this back to the global community of DHIS2 implementers and countries.

> "Yeah, that is because this was the major opportunity we got to share back because we have been using whatever developed by other HISPs so for example, this data importer, the tracker data importer from Uganda, it's app that we have been using more frequently and also this HISP Tanzania DHIS2 touch application. Initially, the Android capture was not supporting aggregate data and we use the touch application developed by HISP Tanzania."

HISP Sri Lanka also felt compelled to share the meta-data configuration of their COVID-19 surveillance system.

> "So we felt I come in rather than people trying to do something, I mean, in reinventing the wheel, we can just share it. So it is an opportunity for us to share back to the community. And it's good for them to waste time on doing it over and over again. Now you can they can just use what they have developed. So I think that was the reason why we felt like sharing."

HISP Sri Lanka's meta-data configuration for port-of-entry screening was shared with the UiO which then shared it immediately through webinars, the DHIS2 website, and the DHIS2 community of practice with dozens of other countries. This happened in late January almost a full month before the initial WHO interim technical guidelines were published on 28 February 2020. Sri Lanka's port-of-entry meta-data configuration became the initial point of reference for fifty-five countries using DHIS2 for COVID-19 Surveillance as of December 2020.

On January 28, 2021 Sri Lanka launched a further expansion of their COVID-19 surveillance system to track the entire Sri Lankan population through vaccination. This further expansion includes an electronic immunization registry, vaccine stock monitoring, dashboards, and custom features for vaccination certificates.

## 5.  ANALYSING DIGITAL RESILIENCE: PRECONDITIONS FOR BOUNCE FORWARD

In our analysis we identified five preconditions that establish a foundation for a state of resilience that produces an improvement or bounce forward of the information system as opposed that is necessary for a return to a status quo in the event of an exogenous shock like COVID-19.  We use the term preconditions instead of following the Heeks and Ospina approach of identifying "enabling attributes" because we feel it more accurately and practically describe our findings in that these conditions have long been pre-established in the country with intention from both the country and the platform owner (UiO). Conditionality also denotes a level of dependency that we also observed between the resilience attributes and the preconditions.

| **Preconditions** | **Sri Lanka** | **Sierra Leone** |
|---|---|---|
| Local Expertise | <ul><li>Existing ICT department with the ability to host the DHIS2 instance.</li><li>HISP Sri Lanka and Medical informatics doctors available to respond.</li><li>Database configured and new apps developed rapidly</li></ul> | <ul><li>Existing eCBS servers running.</li><li>Existing DHIS2 technical experts in place from Ebola response</li><li>COVID-19 added to existing eCBS with minimal effort</li></ul> |
| Platform configurability | <ul><li>HISP Sri Lanka and Medical informatics doctors available to respond.</li><li>Database configured and new apps developed rapidly</li><li>Hackaton and use of core DHIS2 developers at UiO</li></ul> | <ul><li>Existing DHIS2 technical experts in place from Ebola response</li><li>COVID-19 added to existing eCBS with minimal effort</li><li>External DHIS2 technical and disease surveillance experts able to consult on configuration and scaling of eCBS</li></ul> |
| Local platform infrastructure | <ul><li>Existing ICT department with the ability to host the DHIS2 instance.</li><li>ICT Department able to work with MoH for first time</li><li>Key decision makers across multiple departments meeting quickly to approve the system</li></ul> | <ul><li>Existing eCBS surveys running.</li><li>Previous experience from the Ebola response to galvanize actors around system</li><li>Already instutionalized eCBS</li></ul> |
| Local autonomy and ownership | <ul><li>Database configured and new apps developed rapidly</li></ul> | <ul><li>COVID-19 added to existing eCBS with minimal effort</li></ul> |
| | • Key decision makers across multiple departments meeting quickly to approve the system<br>• Need for rapid establishment of port-of-entry screening | • Already instutionalized eCBS<br>• Build on existing system instead of establishing new |
| Distributed Training | • Distributed informatics doctors able to quickly train users<br>• Clinical staff in the field already familiar with DHIS2 | • ToT model from district level to facility<br>• Facility and district medical staff already trained on DHIS2 |

Table 1: Data Table

*Local expertise's* are held by several individuals in the country on adapting and configuring DHIS2 to multiple use-cases or requirements. They are incredibly proficient with all DHIS2 functionality, data models, and components. These individuals are then able to quickly translate disparate requirements from the field into the DHIS2 data model and suite of features and applications. In the case of Sierra Leone, we saw a small group of system architects able to quickly adapt the existing disease surveillance system to also cover COVID-19, and likewise in the case of Sri Lanka, we saw another small team of DHIS2 expert implementers able to quickly capture field requirements and configure a DHIS2 instance that was able to rapidly deploy for port-of-entry screening and patient tracking.

*Platform configurability* was observed to have key three aspects. The first two are that the DHIS2 platform is open source and application based. These two facts, working in concert, showed that new features and functionalities can be rapidly developed by local teams to address urgent situations that are not adequately covered by the generic applications developed by the UiO. We see in the case of Sri Lanka that by holding a hack-a-ton they were able to develop new applications that were necessary for COVID-19 surveillance. As was the case in Sierra Leone where the system administrators were able to expand and adapt existing applications to also cover COVID-19 case registration without the need for complex programming or application development.

*Local platform infrastructure* was observed to be critical. The governments of both case studies were not dependent on licenses, third-party hardware or software vendors, or international platform providers for system hosting or support. The countries have complete access and ownership of their server infrastructure and all the features and functionalities of their DHIS2 instances. This degree of country ownership allowed local DHIS2 system administrators to quickly make changes, update versions, and deploy new features/apps to their DHIS2 instances.

*Local autonomy* is regarding the management and implementation of the information without global support or dependencies on resources outside of the country. This country level ownership means that in our case studies the countries were able to manage and implement the information system as necessary and rapidly as required. They were able to directly respond to the swiftly evolving situation on the ground without immediate concern for additional global resources or oversight. In both cases there was a strong sense of ownership of the system, a mentality of self-reliance and pride.

*Distributed Training* was a key precondition in each case-study. The development, implementation and scale of their respective COVID-19 information systems was undoubtedly hastened by the countries having already trained DHIS2 end-user capacity at each level. The long-term investments that the countries had placed in training up cadres of technical experts resulted in a pool of distributed health officers in both countries that could rapidly utilize the new COVID-19 tools and requirements. It also meant that they were quickly able to cascade training down to less proficient users at the lowest levels (facilities, testing centres, quarantine centres, or ports-of-entry).

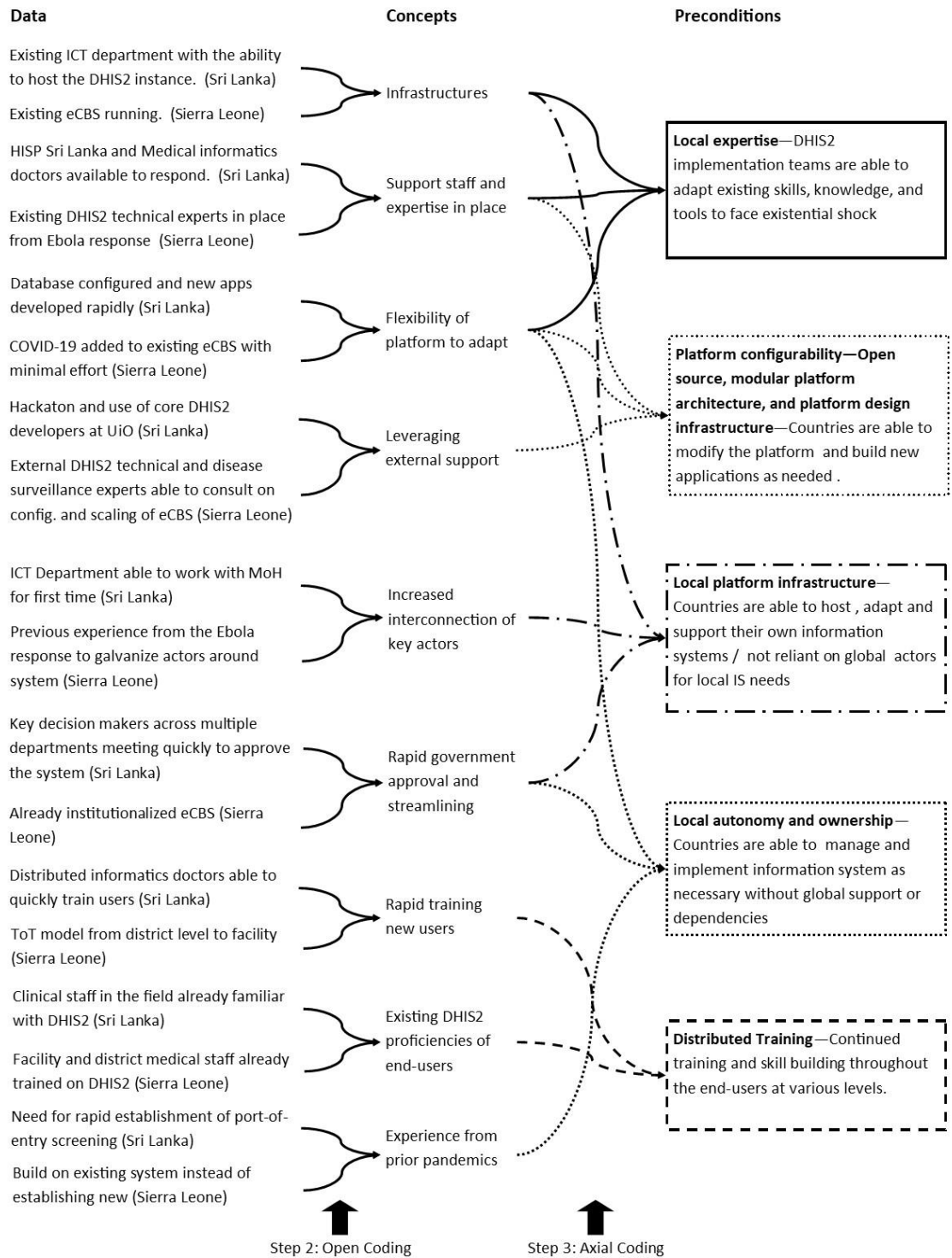

Figure 1: Data coding to preconditions.

The final output of our analysis is the development of a dynamic *bounce-forward resilience model* that illustrates how these identified pre-conditions can lay a foundation for resilience attributes that when exposed to an exogenous shock can result in a bounce forward. Our model (figure 2) shows that there is a complex and interconnected relationship between the preconditions and the resilience attributes. In our data analysis we were not able to reduce a few-to-one or one-to-one relationship between the preconditions and the resilience attributes. We found that multiple preconditions may feed into a single resilience attribute and, while each precondition is unique in its properties, the effects generated by some can easily be seen to touch multiple. However, there are three dependencies that we did appreciate.

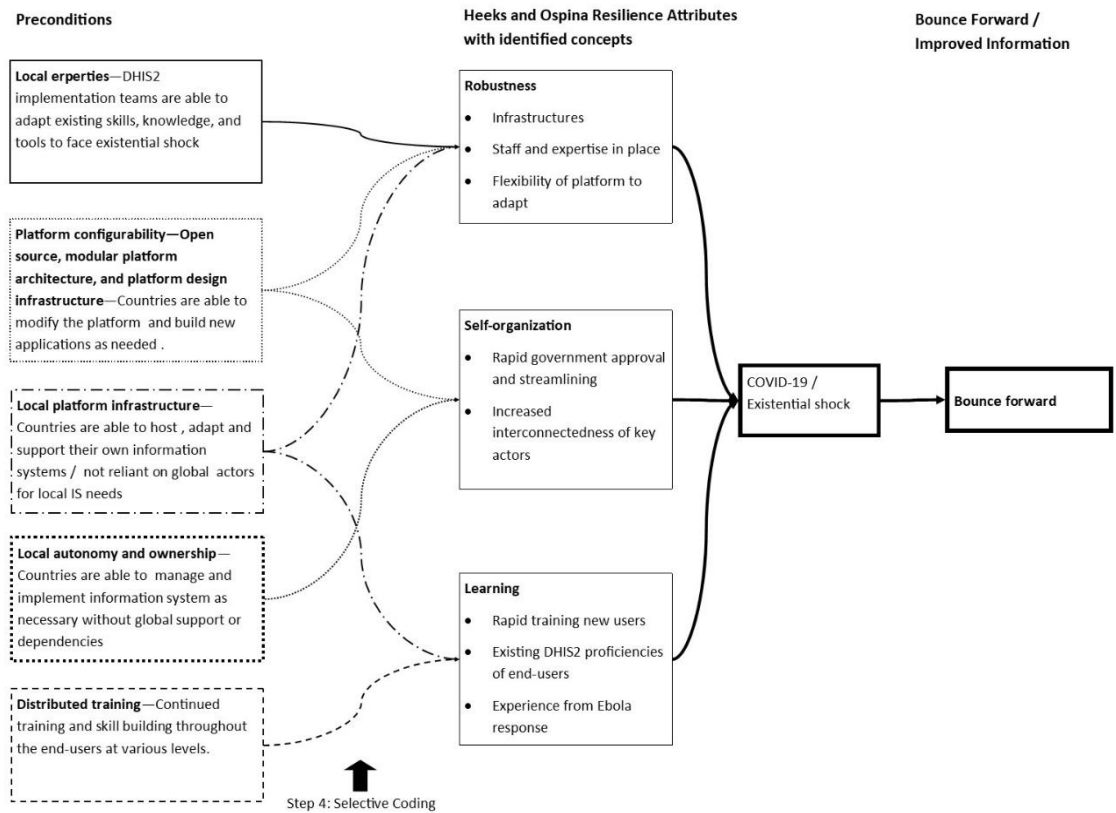

Figure 2: Bounce-forward Resilience Model

*Bounce forward* outcomes were observed in both case-studies, and indeed the goal of this research is to identify new resilience models that incorporate identified causal preconditions to this outcome. In the case of Sierra Leone, the utility of this is already being put to the test with an emerging Ebola outbreak, and in the case of Sri Lanka, it is being further expanded to cover COVID-19 vaccinations. In both cases we observed a swift galvanizing around pre-existing solutions and technologies.

## 6.    IMPLICATIONS AND CONCLUSION

In this research we employ the Heeks and Ospina foundational resilience attributes as a framing for a new, dynamic model that illustrates concrete, operationalizable preconditions. These preconditions form the foundation of a potential for the information system to adapt and bounce forward from a crisis. Heeks and Ospina's own data showed that operationalization of the attributes and markers skewed more toward a bounce-back or a return to normalcy from a crisis. Our research answers the call for the identification of real system properties that enable a long-term change and adaptations. It partially fills the gap in the ICT4D literature for a resilience perspective that is transformative and dynamic [5].

Our model shows that a state of resilience that can harness a crisis for growth is complex with many preconditions and dependencies. It is not so easily distilled down to a few mechanisms, and indeed upon additional observation additional attributes and preconditions would probably be identified. We attempted to focus our preconditions on elements that the country had control over, but certainly looking more broadly it can be appreciated that some preconditions may not be in the country's control.

In our findings, resilience is inherent in the broader system through the defined preconditions. In a time of stability this equates to the ability of the information system to handle day-to-day changes. In the event of an exogenous shock, the ensuing crisis in our cases was not so much the classical shocks of a pandemic, e.g. lack of human resources, overwhelmed health services, economic hardship, unpreparedness, etc. The shocks in our cases to the information system were the need for rapid adaptation, rapid scale, increased investment, intense focus, and the need for innovation. The resultant bounce forward outcomes of these cases point to the fact that, while in a time of stability the countries were at a steady state of resilience, in actuality the countries in time of stress were operating at a higher level of resilience that enabled the bounce-forward outcomes observed. We denote *bounce-forward resilience* as being distinct in terms of its pre-conditions and outcomes in the event of an exogenous shock like COVID-19. Bounce-forward resilient information systems can propel beyond their pre-shock state, develop, and expand in the event of an external shock as opposed to a resilient information system which would maintain or return to a status quo, seemingly unaffected by the shock.

A major limitation of this research is that this model has not yet been able to be applied to other case-study to test its generalizability. Further research must be conducted to address this gap.

## CITATIONS